\documentclass[aps,pre,twocolumn,superscriptaddress,showpacs]{revtex4-1}
\usepackage{epsfig}
\usepackage{amsmath,amssymb,bm}
\usepackage{graphics}
\usepackage{epsfig}
\usepackage{graphicx}
\usepackage{verbatim}
\begin{document}
\title{
%%$e$-
%%Giant viable components of directed interdependent networks
Critical exponents of the explosive percolation transition
}
\author{R.~A.~da~Costa}
\affiliation{Departamento de F{\'\i}sica da Universidade de Aveiro $\&$ I3N, Campus Universit\'ario de Santiago, 3810-193 Aveiro, Portugal}
\author{S.~N. Dorogovtsev}
\affiliation{Departamento de F{\'\i}sica da Universidade de Aveiro $\&$ I3N, Campus Universit\'ario de Santiago, 3810-193 Aveiro, Portugal}
\affiliation{A.F. Ioffe Physico-Technical Institute, 194021 St. Petersburg, Russia}
\author{A.~V. Goltsev}
\affiliation{Departamento de F{\'\i}sica da Universidade de Aveiro $\&$ I3N, Campus Universit\'ario de Santiago, 3810-193 Aveiro, Portugal}
\affiliation{A.F. Ioffe Physico-Technical Institute, 194021 St. Petersburg, Russia}
\author{J.~F.~F. Mendes}
\affiliation{Departamento de F{\'\i}sica da Universidade de Aveiro $\&$ I3N, Campus Universit\'ario de Santiago, 3810-193 Aveiro, Portugal}
%%
%%\date{}
%%
%%\maketitle
\begin{abstract}
In a new type of percolation phase transition, which was observed in a set of non-equilibrium models, each new connection between vertices is chosen 
%%n unusual phase transition was recently discovered in a bunch of evolving networks models in which each new connection between nodes is selected from several 
from a number of possibilities by an Achlioptas-like 
%%Metropolis-like 
%%optimization 
algorithm. 
This causes preferential merging of small 
%%smaller finite 
components and delays the emergence of the percolation cluster. 
First simulations led to a conclusion that a percolation cluster in this irreversible process is born discontinuously, by a discontinuous phase transition, which results in the term ``explosive percolation transition''. We have shown that this transition is actually continuous (second-order) though with an anomalously small critical exponent of the percolation cluster. Here we propose an efficient numerical method enabling us to find the critical exponents and other characteristics of this second order transition for a representative set of explosive percolation models 
%%, in which merging clusters are selected 
with different number of choices. 
The method is based on gluing together the numerical solutions of evolution equations for the cluster size distribution and power-law asymptotics. 
For each of the models, with high precision, we obtain critical exponents 
%%and amplitudes, 
and the critical point.     
\end{abstract}
\maketitle
%%
%%

%%64.60.ah, 05.40.?a, 64.60.F?

%%%%%%%%%%%%%%%
%%%%%%%%%%
%%%%%%%%%%%%%%
%%%%%%%%%%%

\section{Introduction}
%\section{Explosive Percolation Problem}
\label{sec:1}

A phase transition in traditional percolation problems is well known to be continuous, i.e., the order parameter $S$ emerges continuously,
without a jump at the critical point 
%%Percolation problems are well known to show a continuous phase transition which in essence is the birth of a percolation cluster (a giant connected component, as it is called in graph theory) in a system 
\cite{Stauffer:sa-book94,Stauffer:s79}. Above the percolation threshold, a giant connected component (percolation cluster) is present in a system, while below that point all connected components (clusters) are finite. 
This transition is observed for percolation on lattices and on various networks \cite{Dorogovtsev:dm02,Dorogovtsev:dm-book03,Dorogovtsev:dgm08,Dorogovtsev:d-book10}. 
For lattices, the percolation transition is of the second order with the $\beta$ exponent of the order parameter (the size of the percolation cluster) smaller or equal to $1$ ($\beta=1$ in the mean-field regime, i.e., at or above the upper critical dimension of a system). For highly heterogeneous, e.g., scale-free,  networks, the exponent $\beta$ may be above $1$, which corresponds to an order of this transition higher than second \cite{Dorogovtsev:dgm08}. 
The simplest model of percolation (classical random graph) is formulated in the following way. Starting from a large number $N$ of isolated vertices, at each step we choose at random a pair of vertices and interconnect them. When the relative number of links $t=L/N$ in this graph exceeds
the threshold $t_c=1/2$, the graph 
%%includes 
has a percolation cluster containing a finite fraction
$S$ 
%%=N_{{\rm GC}}/N$ 
of all vertices. 
%%This process can be reversed and so the percolation transition in this system is actually equilibrium. 
This is an equilibrium transition since this process can be reversed. 
%%The continuity of the transition means that the order parameter $S$ emerges continuously, without a jump at the critical point. 
%%In particular, for 
For the classical random graph model, in the asymptotic relation
%%described above, 
$S \propto (t-t_c)^\beta$ near the percolation threshold, the critical exponent $\beta$ is $1$. 
%%the relative size of the giant connected component $S$ is proportional to the difference $(t-t_c)$ in the neighbourhood of the percolation threshold, that is the critical exponent $\beta$ defined as $S \propto (t-t_c)^\beta$, is $1$. 
In the neighborhood of the continuous phase transition, scaling behavior takes place. In particular, at the critical point, the cluster size distribution $n(s)$ (which is the probability that a finite cluster contains $s$ vertices), asymptotically, decays as a power law, $n(s)\sim s^{-\tau}$, where the critical exponent $\tau$ is $5/2$ for the classical random graphs. 

%%The continuity of percolation phase transitions was generally accepted until recently. This common 
The common understanding of the percolation phase transition as continuous  
was shaken by the study \cite{Achlioptas:ads09} reporting  
a discontinuous 
percolation phase transition  
in models where each new edge was selected from several possibilities by a Metropolis-like local optimization algorithm (e.g., of two 
%%attempted 
candidate connections, the edge joining two smallest clusters was chosen). 
The  suggested discontinuity resulted in the new term, namely ``explosive percolation". 
%%The suggested discontinuity produced the impressive-sounding term, explosive percolation. 
This observation was confirmed in a number of subsequent works based on simulations, including Refs.~\cite{Araujo:aaz11,Cho:ckp09,Cho:ckk10,Radicchi:rf09,Ziff:z09,Ziff:z10}. 
%%One should note, however, that all these studies, including the first one were based on simulations. 
In our work \cite{daCosta:ddgm10}, we showed that the conclusions for the local optimization based models obtained from these simulations 
%%of our predecessors 
were incorrect, and the so-called explosive percolation transition is actually continuous for infinite systems. We explained that the exponent $\beta$ of this transition is surprisingly small, which makes the observation of the continuous transition 
%%virtually impossible 
in simulations of realistic size systems 
%%of these systems 
virtually impossible. 
The critical singularity with a small $\beta$ is perceived as a discontinuity for simulated systems. 
The continuity of the explosive percolation transition was afterwards 
%%proved mathematically 
confirmed by mathematicians \cite{Riordan:rw11} and was observed in Ref.~\cite{Nagler:nlt11,Lee:lkp11,Grassberger:gcb11,Qian:qhm12} for other models.  

To describe quantitatively the explosive percolation transition, in our work \cite{daCosta:ddgm10} we 
showed that the problem can be formulated as a specific aggregation process. 
%%reduced the problem to a specific aggregation process and solved numerically $10^6$ evolution equations 
%%%%of the Smoluchovsky equation type 
%%for the cluster size distribution, which corresponds to the range of cluster sizes $s \leq 10^6$. 
The evolution equations for the explosive percolation problems resemble the Smoluchowski equation \cite{Smoluchowski:s1915}, which enables us to use traditional numerical algorithms \cite{Leyvraz:lt81}. 
The system of equations was conveniently organized in such a way that we could solve them one by one, sequentially. 
In this way we succeeded in solving numerically $10^6$ evolution equations 
%%%%of the Smoluchovsky equation type 
for the cluster size distribution, which corresponds to the range of cluster sizes $s \leq 10^6$. 
Nonetheless, these direct numerical calculations were so computationally demanding that proceeding in this way we could not further improve the precision of our results including the critical exponents and amplitudes and the explosive percolation threshold position. In the present work we present a new numerical approach to this problem.  
We demonstrate how to find characteristics of the explosive percolation transition 
%%even 
with higher precision by implementing an 
%%essentially more 
efficient method, without solving a 
%%so 
large array of evolution equations.

%%%%%%%%%%%%%%%
%%%%%%%%%%
%%%%%%%%%%%%%%
%%%%%%%%%%%

\section{The model}
\label{sec:2}

\begin{figure}[b]
\begin{center}
\includegraphics[scale=.8]{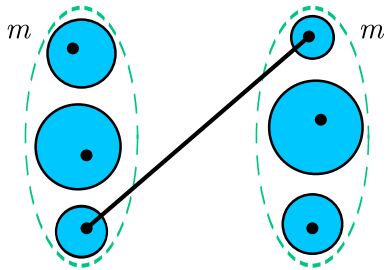}\end{center}
\caption{
%%Schematic representation of the 
Explosive percolation model rules. At each step, two samples of $m$ vertices are selected 
%%uniformly 
at random. In each of the samples, 
%%the 
a vertex belonging to the smallest (of $m$) cluster is chosen, and a new edge connecting these two vertices is added to the system. 
%%This can be treated as a specific aggregation process, in which 
%%In other words, two sets of clusters are chosen with probability proportional to their sizes and two smallest clusters, taken from each of the sets, merge together. 
}
\label{f1}       % Give a unique label
\end{figure}

Let us employ the version of the explosive percolation process first considered in Ref.~\cite{daCosta:ddgm10}. This model belongs to the same class as the original so-called Achlioptas process simulated in Ref.~\cite{Achlioptas:ads09} and in simulations it produces the same seemingly discontinuous phase transition. Moreover, the exponent $\beta$ in our model turns out to be even smaller than in the process from Ref.~\cite{Achlioptas:ads09}. 
Our representative and actually elegant model naturally generalizes the classical random graph model of ordinary percolation (see Sec.~\ref{sec:1}) and allows for analytical and numerical treatment. 
The process is defined in the following way. 
We start from 
%%a given set of clusters, 
an arbitrary initial configuration, for example, from a large number $N$ of isolated vertices, and at each step, we 
%%make the following, see Fig.~\ref{f1}. We 
select uniformly at random $m$ vertices ($m$ vertex sample) and choose that of them which is inside the smallest of the clusters to which these vertices belong.   
%%.   Each of these nodes belong to some cluster; the number of these clusters is $m$ or less, if some of the nodes belong to the same clusters. Of the $m$ nodes and choose that of them which belongs to the smallest  
Then we again select $m$ vertices and choose that of them belonging to the smallest cluster, and, finally add an edge connecting the two vertices selected in this way, see Fig.~\ref{f1}. 
In other words, at each step, two sets of clusters are chosen with probability proportional to their sizes and two smallest clusters, taken from each of the sets, merge together. If $m=1$, we recover the classical random graph model. Here we consider the cases of $m=2$, $3$, and $4$. 
In the limiting case of $m\to \infty$, the giant cluster discontinuously emerges at the point $t=1$ where evolution ends. At this point, the relative size of the giant cluster jumps from $0$ to $1$. Similarly to one-dimensional percolation, this cannot be regarded as a discontinuous transition, since the position of the jump coincides with the end of evolution. 
%%occurs at the border of the complete $t$ interval.

This model can be treated as an aggregation process, in which clusters selected by our rules merge progressively. A complete description of this process is provided by the evolving distribution $P(s,t)=sn(s)/\langle s \rangle$, which is the probability that a uniformly randomly chosen vertex belongs to a cluster of size $s$ at time $t$. Here $n(s)$ is the size distribution of clusters and $\langle s \rangle$ is the average size of clusters (including the percolation cluster). 
Let us introduce the probability $Q(s,t)$ that 
%%the 
a vertex chosen by our rules from an $m$ vertex sample belongs to a cluster of size $s$.
%%Another important characteristic is the probability $Q(s,t)$ that the node chosen by the rules formulated above from an $m$ node sample belongs to a cluster of size $s$. In other words, t
This is the size distribution of merging clusters. 
%%selected for merging by our rules. 
Using 
%%basic 
formulas of probability theory \cite{Feller:f68}, 
%%one can 
we express the distribution $Q(s,t)$ in terms of the distribution $P(s,t)$, 
\begin{equation}
Q(s,t)\! =\! \!\left[ \sum_{u = s}^\infty P(u,t){+}S(t) \!\right]^m\!\!\!\! - \left[\!\sum_{\ u = s+1}^\infty \!\!\!P(u,t){+}S(t) \!\right]^m 
\!\!\!\!,
\label{e10}
\end{equation}
which is the basic formula of extreme value statistics. 
Here $\sum_{u = s}^\infty P(u)+S$ is the cumulative distribution. 
This is 
the probability that a uniformly randomly chosen vertex belongs to a cluster of size $u\geq s$ including 
%%the probability that a vertex belongs to 
the giant component. 
Using the normalization condition $\sum_{u = 1}^\infty P(u)+S=1$, we obtain  
%%namely  
%%
\begin{equation}
Q(s,t)=\left[ 1-\sum_{u = 1}^{s-1} P(u,t) \right] ^{m}\!\!\!\! - \left[1- \sum_{u = 1}^{s} P(u,t) \right]^{m} 
.
\label{Q_def}
\end{equation}
%%
%%which is the basic formula of extreme value statistics. 
Note also the relation $\sum_{u = 1}^\infty Q(u)+S^m=1$.
The evolution equations for the distributions $P(s,t)$ and $Q(s,t)$ describing this aggregation process in the infinite system ($N \to \infty$) 
%%under consideration 
%%look as follows: 
have the form
\begin{equation}
\frac{\partial P(s,t)}{\partial t}= s\sum_{u+v=s} Q(u,t) Q(v,t) - 2sQ(s,t)
.
\label{master_eq}
\end{equation}
In the case of $m=1$, $Q(s,t)$ in Eq.~(\ref{master_eq}) should be replaced by the distribution $P(s,t)$, and we arrive at well-known master equations for standard percolation, which can be 
%%easily 
solved explicitly \cite{Krapivsky:krb10}. 
%%by implementing a standard generating function technique. 
This cannot be done 
%%in general, for arbitrary 
for $m>1$ because in this case the right-hand side of Eq.~(\ref{master_eq}) is 
%%already 
not bilinear due to relation (\ref{Q_def}) and cannot be treated by a generating function technique.  
Because of this 
%%principal 
nonlinearity, the case of $m\geq2$ is far more difficult than that of $m=1$. 
So we will analyze 
%%We should analyse difficult 
Eq.~(\ref{master_eq}) numerically, taking into account the relation (\ref{Q_def}) and a given initial distribution $P(s,0)$. 
%%and obtain 
In our work \cite{daCosta:ddgm10} we showed that the solution of this equation has a power-law asymptotics 
\begin{equation}
P(s,t_c) \cong f(0) s^{1-\tau} 
\label{e3}
\end{equation}
at the critical point, where the exponent $\tau$ slightly exceeds $2$, which indicates a continuous phase transition. 
Note that there exist phase transitions combining discontinuity and critical singularity which show a power-law distribution $P(s)\sim s^{-3/2}$ at the critical point \cite{Dorogovtsev:dgm06}. In these so-called hybrid transitions, the critical singularity is present when one approaches the critical point only from one side of the transition, namely from the ordered phase. This (the value $\tau=5/2$ and the asymmetry of the hybrid transition) differs sharply from the situation considered in this work. Nonetheless, our method is in principle applicable to hybrid phase transitions as well.   
The factor $f(0)$ in Eq.~(\ref{e3}) is a critical amplitude which 
%%coincides with 
equals the value of the scaling function $f(x)$ at $x=0$. 
At $t\neq t_c$,  the size distribution of clusters decreases more rapidly than any power law. 
Near $t_c$ we have  
$P(s,t)=s^{1-\tau}f\left(s|t-t_c|^{1/\sigma}\right)$, where $\tau$ and $\sigma$ are critical exponents. 
While the values of the critical exponents are independent of initial conditions, 
%%cluster size distributions,  
%%(initial distributions should decay sufficiently rapidly to produce a nonzero critical point, $t_c>0$), 
the form of the scaling function $f(x)$, and so the critical amplitude $f(0)$, depends on the initial distribution $P(s,0)$. 
Note that initial distributions should decay sufficiently rapidly, faster than a power law with exponent 
$-2m/(2m-1)$, to produce a nonzero critical point, $t_c>0$. The role of initial conditions will be considered elsewhere. 
Furthermore, the scaling function above the critical point differs from that below $t_c$  \cite{daCosta:ddgm10}. 
The dimensionality of our system is infinite, i.e., above the upper critical dimension, which guarantees an exact 
%%mean-field theory 
description 
in the framework of a mean-field approach. Exact Eqs.~(\ref{Q_def}) and (\ref{master_eq}) provide this description. 
As it should be above the upper critical dimension, one can express any critical exponent in this problem in terms of one of them. So we need to find one critical exponent. 
%%In our work \cite{daCosta:ddgm10}, we have derived the relations between the critical exponents of this phase transition in the particular case of $m=2$. In a similar way, 
Similarly to Ref.~\cite{daCosta:ddgm10}, we find the following relation between the critical exponents $\tau$ and $\beta$ for arbitrary $m$:
\begin{equation}
\beta =\frac{\tau-2}{1-(2m-1)(\tau-2)}
%%.
\label{e4}
\end{equation}
and the expression for the upper critical dimension 
\begin{equation}
d_{\text{uc}} =2+4m\beta 
%%.
\label{e4a}
\end{equation}
above which a mean-field description is exact. 
The detailed derivation of these relations and the complete scaling theory of the explosive percolation transitions will be presented elsewhere. 
In Eq.~(\ref{e4a}) we exploited the fact that the upper critical dimension can be always expressed in terms of the mean-field theory critical exponents. 
Note that our models, in the present form, are defined in infinite dimensions, and, formally speaking, one cannot directly implement them in lower dimensions. 
%%There were considered a bunch of explosive percolation systems on finite dimensional lattices \cite{Ziff:z09}. 
%%Here $d_{\text{uc}}$ is 
%%a number having the meaning of the upper critical dimension of 
%%actually for the corresponding model of the same universality class of critical behavior (i.e., with the same set of critical exponents) defined on lattices (if this model exists). 
Assume that one can introduce an explosive percolation model of the same universality class of critical behavior (i.e., with the same set of critical exponents) but defined on lattices. Then $d_{\text{uc}}$ is the upper critical dimension of this model. 
Notice that 
%%there were considered 
a bunch of explosive percolation systems on finite-dimensional (mostly two-dimensional) lattices were considered~\cite{Ziff:z09}.

%%We will 
Let us obtain the critical exponent 
%%Thus our main aim is to find one of the critical exponents, say, 
$\tau$ and the critical amplitude $f(0)$ for $m=2$, $3$, and $4$, as well as the critical point (explosive percolation threshold) $t_c$. 
%%For the sake of brevity, here we 
We assume that at the initial moment all vertices are disconnected, that is $P(s,0)=\delta_{s,1}$, where $\delta_{s,1}$ is the Kronecker symbol. This is our initial condition for the evolution equation. 

%%Equation~(\ref{master_eq}) with Eq.~(\ref{Q_def}) substituted is a chain of coupled equations, in which the first equation yields $P(1,t)$, substituting this result into the second equation and solving it yields $P(2,t)$, and so on. 
Substituting Eq.~(\ref{Q_def}) into Eq.~(\ref{master_eq}) we find $P(1,t)$. Substituting this result into the second equation and solving we obtain $P(2,t)$, and so on. 
%%, see the forms of Eqs.~(\ref{Q_def}) and (\ref{master_eq}). 
This procedure enables us to solve numerically the first $s_{\text{max}}$ equations and find $P(s\leq s_{\text{max}},t)$ at any $t$ with any desired precision. 
%%In our work \cite{daCosta:ddgm10} we fulfilled this program, directly solved the evolution equations, and found $P(s\leq 10^6,t)$ including the critical distribution in the particular case of $m=2$. Let us use a more efficient way to obtain the characteristics of the explosive percolation transition.     

%%%%%%%%%%%%%%%%%%%%%%%%%%%%%%
%%%%%%%%%%%%%%%%%%%%%%%%%%%%%%
%%%%%%%%%%%%%%%%%%%%%%%%%%%%%%
%%%%%%%%%%%%%%%%%%%%%%%%%%%%%%

\section{The approach}
%%{The method}
\label{sec:3}

The method we use in this paper is based on the fact that at the critical point, the cluster size distribution $P(s,t_c)$ has the power-law asymptotics (\ref{e3}), whose parameters, $\tau$ and $f(0)$ we do not know yet. 
%%We use 
%%numerical solution provided by 
%%distributions obtained by numerical solution of  
%%We 
%%can 
%%solve numerically 
%%a finite number $s_{0} \leq s_{\text{max}}$ of master equations~(\ref{master_eq}). 
%%, say $s_{0}$, using the sampling rule~
%%taking into account relation (\ref{Q_def}). 
%%This will provide us with $P(s,t)$ for $s\leq s_{0}$ and any time $t$. 
The idea is to glue together the numerical solution $P(s,t)$ and a power-law function at some cluster size $s_0 \leq s_{\text{max}}$ and then analyze the variation of the result with $s_0$. 
%%The numerical solution of the system of $$ master equations allows us to find the $P(s,t)$ for $s<s_{}$ 
Let us assume first that we know precisely the value of the critical point $t_c$, which is actually not the case. 
Then, after finding numerically $P(s,t_c)$ for all $s\leq s_{0}$ and gluing it to a power law at $s_{0}$, we 
%%will 
easily obtain 
%%If we knew the value of the critical point $t_c$, we would 
%%readily obtain 
%%the 
exponent $\tau$ and 
%%the 
critical amplitude $f(0)$. 
%% by sewing together the distribution $P(s\leq s_{0},t_c)$ obtained by numerical solution of the evolution equations and the power-law asymptotics $P(s\geq s_{0},t_c) = f(0)s^{1-\tau}$. 
For that, one 
%%should 
uses two conditions: (i) $P(s_{0},t_c) = f(0)s_{0}^{1-\tau}$ and (ii) the normalization condition, namely, 
\begin{equation}
\sum_{s=1}^\infty P(s,t_c) = \sum_{s < s_{0}} P(s,t_c) + f(0)\sum_{s \geq s_{0}} s^{1-\tau}= 1
.  
%%\nonumber
%\label{eq:norm cond_EP}
\end{equation}
%%
%%Sewing condition (i) 
This condition can be conveniently rewritten in the following form:
%%The normalization condition can be written to take into account two parts separately:
%%%%
%%\begin{equation}
%%1=\sum_{s < s_{0}} P(s,t_c) + f(0)\sum_{s \geq s_{0}} s^{1-\tau}.
%%%%\nonumber
%%%\label{eq:norm cond_EP}
%%\end{equation}
%%%%
%%It is convenient to have an expression with a single unknown variable, which could then be easily calculated. To find such an expression, we now substitute $f(0)$ given by condition (i) into last equation, 
%\begin{align}
%1&=\sum_{s\ < s_{\rm min}} P(s,t_c) + P(s_{0},t_c)s_{0}^{\tau-1}\sum_{s \geq s_{0}} s^{1-\tau}, 
%\nonumber
%\\[5pt]
%&=\sum_{s\ < s_{\rm min}} P(s,t_c) + P(s_{0},t_c)s_{0}^{\tau-1}\left(\zeta(1-\tau) -\sum_{s < s_{0}} s^{1-\tau}
%\label{eq:norm_cond_CEPT}
%\end{align}
%%
\begin{equation}
\hspace{-0.15pt}
1=\sum_{s< s_0}P(s,t_c) + P(s_{0},t_c)s_{0}^{\tau{-}1}\!\!\left[\zeta(\tau{-}1)-\!\sum_{s < s_{0}}\! s^{1-\tau}\!\right]\!,\!\!\!
\label{eq:norm_cond_CEPT}
\end{equation}
where $\zeta(x)=\sum_{s=1}^\infty s^{-x}$ is the Riemann zeta function.
%%The first sum on the right-hand side of this expression is provided by the numerical solution of $s_{0}$ evolution equations, as well as the multiplicand $P(s_{0},t_c)$ of the second term. 
With known $t_c$, we would immediately find critical exponent $\tau$ from this equation, 
leading to the precise value of $\tau$ 
%%and the result be precise 
in the limit $s_{0}\to \infty$. 
%%Therefore, since everything else is known in equation~(\ref{eq:norm_cond_CEPT}), we could now solve this equation for the unknown $\tau$. 
%%This  result would be precise in the limit $s_{0}\to \infty$. 
%%
%%%%%%%%%%%%%%%%%%%%%%%%%%%%%%%%%%%%%%%%%
%%%%%%%%%%%%%%%%%%%%%%%%%%%%%%%%%%%%%%%%%
%%
%%
\begin{figure}[t]
\includegraphics[scale=.33]{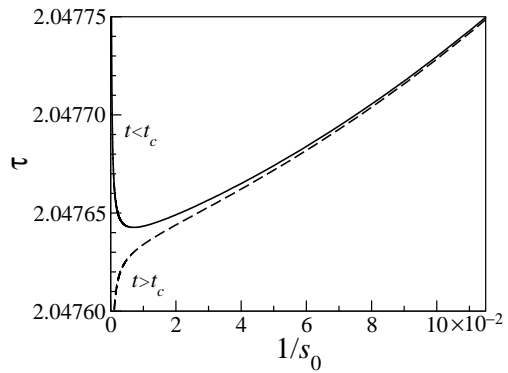}
\caption{
Variation of $\tau$ for $m=2$, calculated from Eq.~(\ref{eq:norm_cond_2_CEPT}), with $1/s_{0}$, where $s_{0}$ is the 
%%sewing 
point at which numerical solution $P(s,t)$ is glued together with a power law. 
%%denotes the maximum cluster size taken into account in the evolution equations for $m=2$. 
In our calculations, $s_{0}$ is in the range up to $10^5$. The solid and dashed curves correspond to two values of $t$, 
%%below and above the transition,  
namely, $t=0.923207<t_c$  and $t=0.923208>t_c$, respectively. 
}
\label{f2}       % Give a unique label
\end{figure}
%%
%%%%%%%%%%%%%%%%%%%%%%%%%%%%%%%%%%%%%%%%%
%%%%%%%%%%%%%%%%%%%%%%%%%%%%%%%%%%%%%%%%%
%%
\begin{figure}[t]
%%\sidecaption
% For example, with the graphicx style use
\begin{center}
\includegraphics[scale=.32]{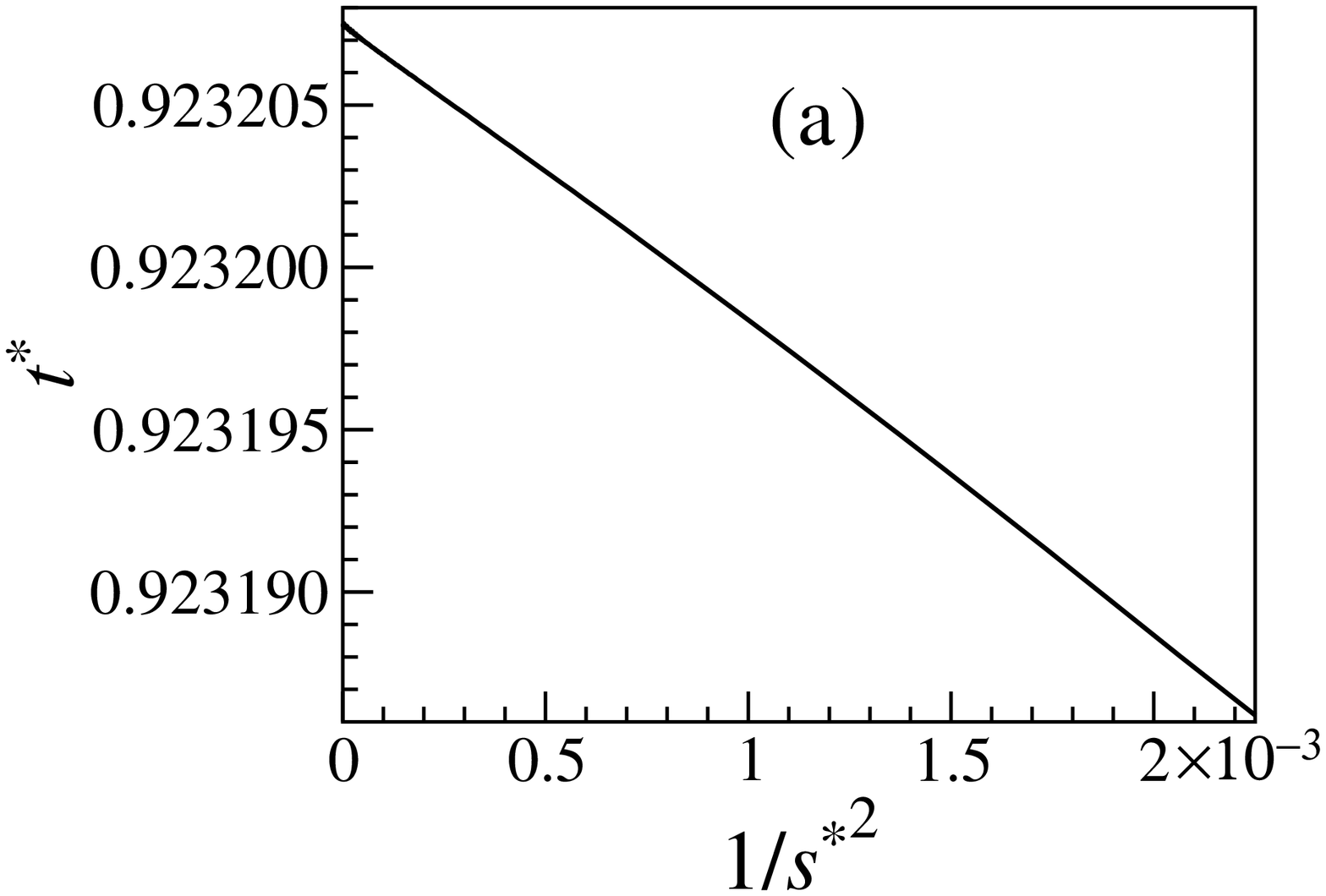}
\\[01pt]
\hspace{0.45pt} \includegraphics[scale=.32]{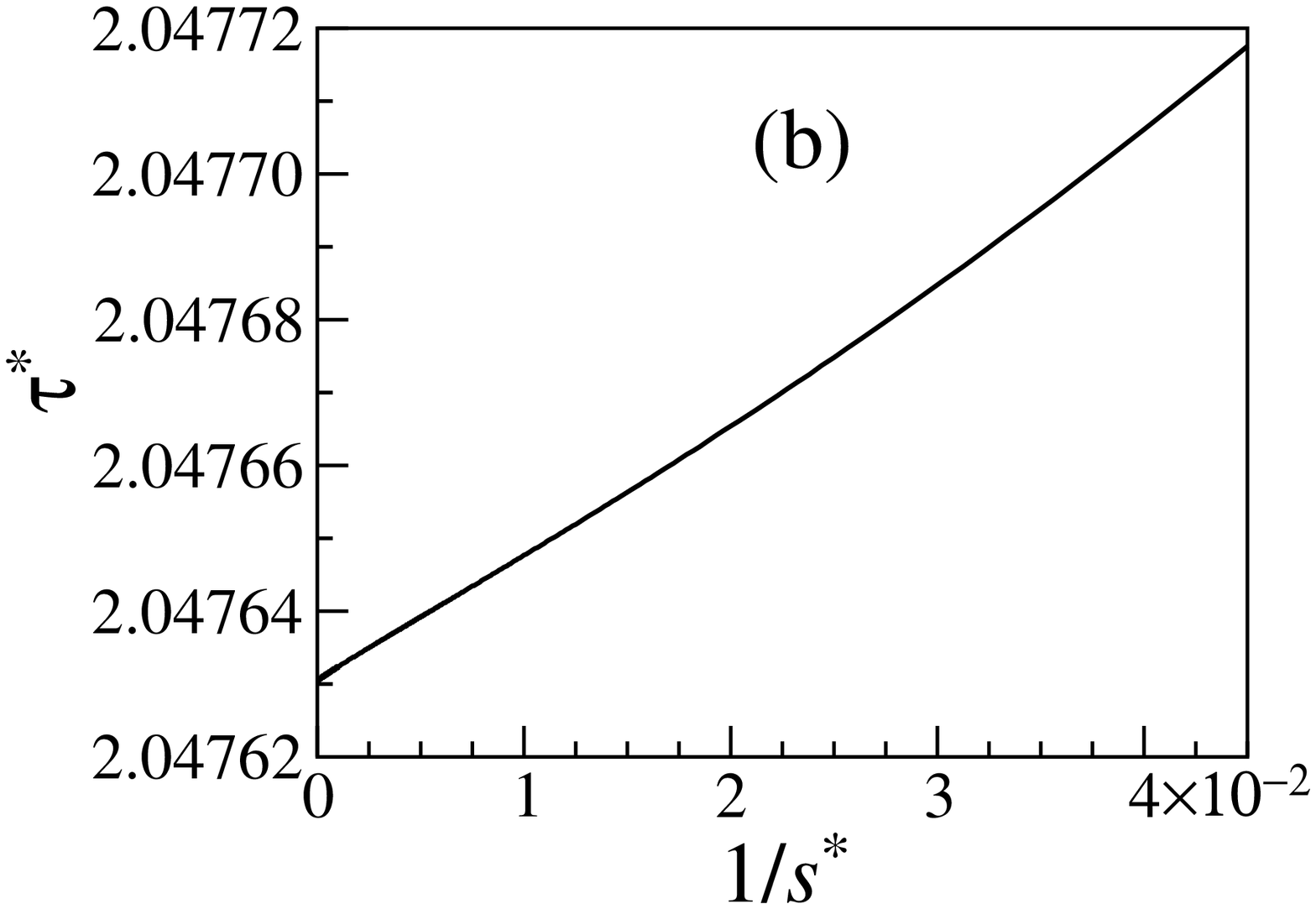}
\\[4pt]   
\hspace{-0.9pt} \includegraphics[scale=.32]{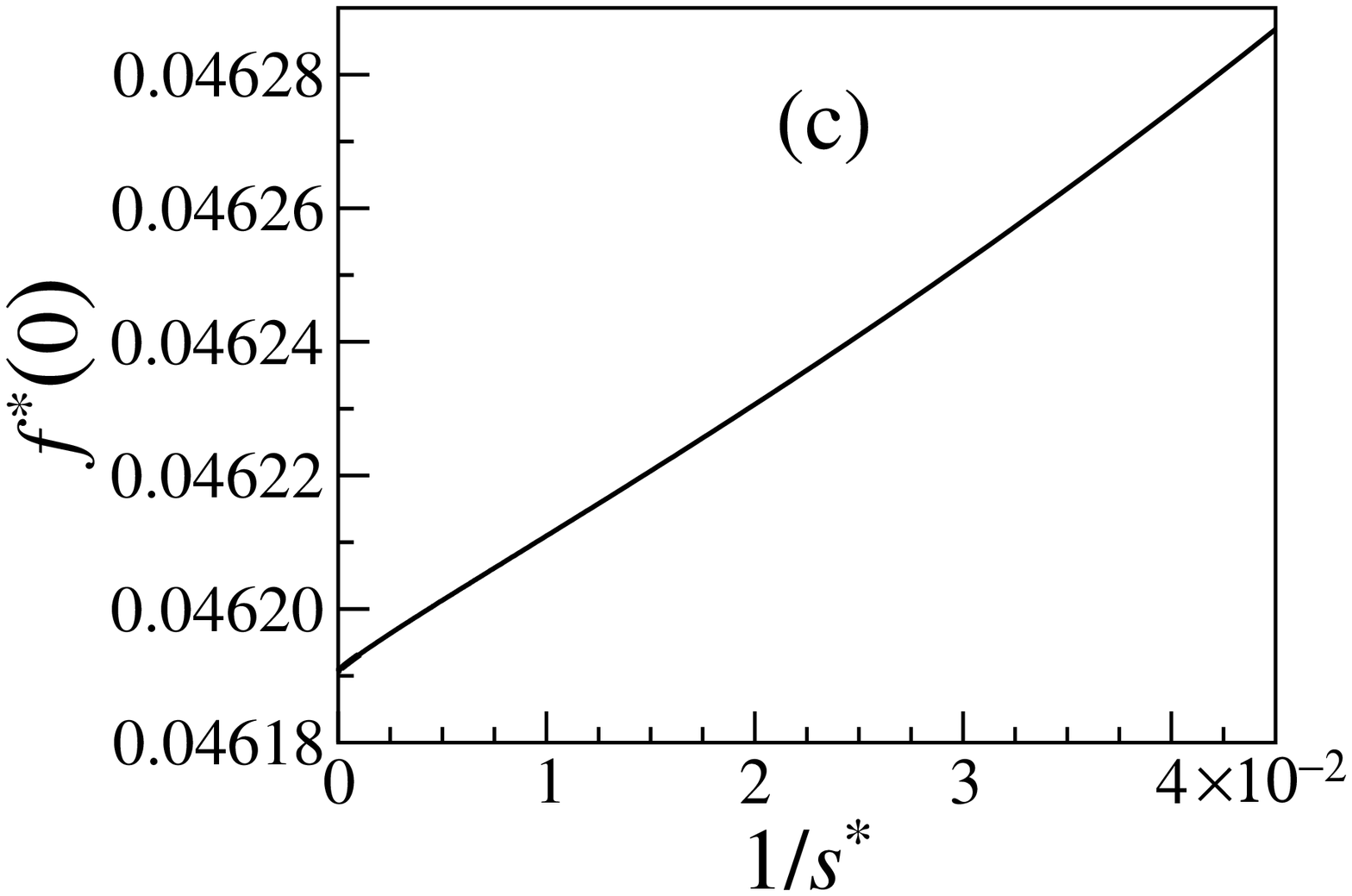}
%%\\[36pt]
\end{center}
%
% If no graphics program available, insert a blank space i.e. use
%\picplace{5cm}{2cm} % Give the correct figure height and width in cm
%
\caption{
Functions 
%%Extrapolation of the curves: 
%%Dependencies 
(a) $t(s^\ast)$ vs. $1/{s^\ast}^2$, and (b) $\tau(s^\ast)$ and (c) $f^*(0,s^\ast)$ vs. $1/{s^\ast}$  (see the text for the definitions of these functions) in the case of $m=2$.  
%%$t^\ast$, 
%%$t$, (b) $\tau^\ast$, and (c) $f^\ast(0)$, corresponding to 
%%%%the position of 
%%the mimima
%%$1/s^\ast$ of the $\tau(1/s_{0},t<t_c)$ 
%%%%$s^\ast$ 
%%of 
%%%%the 
%%$\tau(s_{0},t<t_c)$ 
%%%%curves 
%%%%found from 
%%in Fig.~\ref{f2}, vs. inverse positions of the minima $s^*$ 
%%%%to the precise values $t_c$, $\tau$, and $f(0)$, respectively, at $s^\ast\to\infty$ 
%%in the case of $m=2$. 
%%Here $\tau^* = \text{min}_{s_{0}} \tau(s_{0},t)$, and $s^*$ is the value of $s_{0}$ at which this minimum takes place, $f^\ast(0)$ is calculated from the relation $f^\ast(0) = P(s^*,t){s^\ast}^{\tau^*-1}$. 
In the limit $s^*\to\infty$, the values $t(s^*)$, $\tau^*$, and $f^\ast(0)$ approach the exact values of the critical point $t_c$, the critical exponent $\tau$, and the amplitude $f(0)$, respectively.
These curves are extrapolated to $1/s^*{\to} 0$ to obtain the precise values  $t_c$, $\tau$, and $f(0)$.
}
\label{f3}       % Give a unique label
\end{figure}
%%
%%
%%%%%%%%%%%%%%%%%%%%%%%%%%%%%%%%%%%%%%%%%
%%%%%%%%%%%%%%%%%%%%%%%%%%%%%%%%%%%%%%%%%
%%
The value of $t_c$ however is not known in advance. 
Nonetheless, 
%%Although the value of $t_c$ is not known in advance, 
we can formally perform this procedure at any $t$, inserting the numerical solution of the evolution equations for $s \leq s_{0}$ into the following equations: 
%%, not necessarily equal to $t_c$. In other words, there is no impediment to the application of last expression to the numerical results for $P(s \leq s_{0},t)$ at any $t$:
 %%
\begin{eqnarray}
&&\hspace{38pt}
P(s_{0},t) = f(0)s_{0}^{1-\tau}
%%1=\sum_{s < s_{0}} P(s,t) + f(0)\sum_{s \geq s_{0}} s^{1-\tau}
,
%%\nonumber
\label{90}
\\[5pt]
&&
\hspace{-29pt}
%%P(s_{0},t) = f(0)s_{0}^{1-\tau}
1=\!\sum_{s< s_0} \!P(s,t) + P(s_{0},t)s_{0}^{\tau-1}\!\left[\zeta(\tau{-}1) -\!\sum_{s < s_{0}}\! s^{1-\tau}\right].\!\!\!
\label{eq:norm_cond_2_CEPT}
\end{eqnarray}
From which we can find $\tau$ and $f(0)$ vs. $s_{0}$ for any $t$.
Clearly, the exact value of $\tau$ is given by the last equation only for $t=t_c$ (when $s_{0}\to \infty$). However, we can still use it to find the set of points $\tau(s_{0})$ corresponding to some time $t \neq t_c$. 
%%Therefore we 
The idea is to analyze 
%%variation  
%%the behavior 
%%of solutions $\tau(s_{0},t)$ of Eq.~(\ref{eq:norm_cond_2_CEPT}) with $s_{0}$ 
how the solutions $\tau(s_{0},t)$ of Eq.~(\ref{eq:norm_cond_2_CEPT}) vary with $s_{0}$ for a set of $t$ chosen from a neighborhood of the supposed critical point. 
Below and above the critical point, 
%%there dependencies 
these solutions behave quite differently as $s_{0}$ approaches infinity, and so it will be easy to identify $t_c$. 
This difference is due to the fact that above $t_c$, Eq.~(\ref{eq:norm_cond_2_CEPT}) 
%%does not take into account 
%%ignores the presence of 
neglects 
the giant component, and  the real sum for finite components $\sum_{s=1}^\infty P(s,t>t_c)$ actually equals $1-S<1$.  
%%in the phase with a giant component, $t>t_c$, the normalization 
 
 Figure~\ref{f2} shows two typical curves $\tau$ vs. $1/s_{0}$ for two particular values of $t$, one below and the other above the critical point (respectively, solid and dashed curves)
%%The results obtained at two values of $t$, below and above $t_c$, 
%%are shown in Fig.~\ref{f2} 
for the case of $m=2$. 
%%There, 
%%%%for the sake of brevity, $s_{0}$ is denoted by simply $s$, 
%%the solid curve shows $\tau(s_{0})$ at some $t<t_c$ (normal phase), and the dashed curve shows $\tau(s_{0})$ at some $t>t_c$, which corresponds to a phase with a giant component. 
%%The plot shows how the value of the exponent $\tau$, found by using the sewing procedure (equation~(\ref{eq:norm_cond_2_CEPT})), varies with the inverse maximum cluster size $1/s_{0}$ taken into account in the equations. 
With decreasing $1/s_{0}$, these dependencies approach infinity or $2$ if $t$ is below or above $t_c$, respectively. 
One can show that an infinite exponent $\tau(s_{0}\to\infty)$ corresponds to a rapidly decreasing cluster size distribution in the normal phase ($t<t_c$). 
Indeed, 
%%As explained above, 
%%except at 
apart from the critical point, 
%%for large enough $s$, 
the distribution $P(s,t)$ decays with $s$ in a exponential-like fashion for large $s$. 
Numerical solution of the evolution equations gives a distribution of this kind in the range of $s\leq s_{0}$. 
When we try to glue together this exponential-like function for $s\leq s_{0}$ and a power-law function for $s\geq s_{0}$, we can only get the infinite exponent $\tau$, 
%%of this power law, 
which corresponds to a more rapid decay than any power law. 
%%an exponential decay. 
%%Sewing together this distribution with a power law in the phase without a giant component leads to an exponent $\tau$ reproducing the rapid decay. 
%%That is, for $t<t_c$ 
%%the solution $\tau(s_{0})$ of Eq.~(\ref{eq:norm_cond_2_CEPT}) approaches infinity  as  
%%%%$\tau \to \infty$ when 
%%$1/s_{0}$ goes to $0$.
%
On the other hand, the behavior $\tau(s_{0})\to 2$, when $s_{0} \to \infty$, indicates the failure of the normalization condition 
%%used in the construction of expression~(\ref{eq:norm_cond_2_CEPT}); 
$\sum_{s=1}^\infty P(s,t)=1$, 
which is no longer valid in the phase with a giant component. In this phase, $1-\sum_{s=1}^{s_{0}} P(s,t)$ approaches a nonzero value $S(t)$ as $s_{0}\to\infty$. 
This violation of 
%%the 
normalization to $1$ manifests itself in the divergence of the sum $\sum_{s=s_{0}}^\infty s^{1-\tau}$ on its upper limit at $\tau=2$. 
%%%%As a result, the second term on the right-hand side of Eq.~(\ref{eq:norm_cond_2_CEPT}) should tend to a finite constant
%%%%the attempt to find an exponent $\tau$, such that $P(s_{0},t)s_{0}^{\tau-1}\sum_{s=s_{0}}^\infty s^{1-\tau}=S(t)$, 
%%%%when $s_{0}\to\infty$. 
%%This is possible only if $\tau$ tends to $2$ (if $\tau\leq 2$, this term would diverge). 
%%, leads $\tau\to 2$ (notice that, for normalization reasons, an exponent $\tau\leq 2$ is impossible).

%in this indicates the failure of the power-law form, which we assumed, to describe correctly the distribution of clusters in the phase with a giant component.   
If 
%%we guess $t_c$ properly, 
$t=t_c$, the curve $\tau(s_{0},t)$ 
%%will lead 
leads to the precise $\tau$ as $1/s_{0} \to 0$. Otherwise, the curves run away from that point, which is 
%%precisely 
the behavior demonstrated by the solid and dashed curves in Fig.~\ref{f2}. To find the precise values of $t_c$, $\tau$, and $f(0)$ we inspected a set of $t$. 
For each of these $t$, we find the minimum of $\tau(s_{0})$, see Fig.~\ref{f2}. 
Namely, we find the value of this minimum $\tau^* = \text{min}_{s_{0}} \tau(s_{0},t)$ and the value $s^*$ of $s_{0}$ at which it takes place.  
We also find the value $f^\ast(0)$ corresponding to this minimum, 
$f^\ast(0) = P(s^*,t){s^\ast}^{\tau^*-1}$. 
In particular, these results provide the dependence of the position of the minima $s^*$ on $t$. 
%%It turns out however that instead of considering the function $s^*(t)$, it is more convenient to analyze the transformed inverse function $t$ vs. $1/{s^*}^2$. 
Instead of 
%%considering the function 
$s^*(t)$, it is 
%%more 
convenient to 
%%we 
analyze the 
%%transformed 
inverse function $t$ vs. $1/{s^*}^2$, since  we are interested in the limit $s^*\to\infty$, and the function $t(1/{s^*}^2)$ is close to linear, see Fig.~\ref{f3}(a). 
%%This gives a function $t(1/{s^*}^2)$, see Fig.~\ref{f3}(a).  
In 
%%the
this limit, $s^*\to\infty$, the values $t(s^*)$, $\tau^*$, and $f^\ast(0)$ tend to the exact values of the critical point $t_c$, the critical exponent $\tau$, and the amplitude $f(0)$, respectively. 
This figure demonstrates that the curve $t$ vs. $1/{s^*}^2$ approaches $t_c$ 
almost  
linearly. 
%%with $1/{s^{\ast}}^{ 2}$ 
Similarly, we plot $\tau^*$ and $f^\ast(0)$ vs. $1/s^*$, see Figs.~\ref{f3}(b) and \ref{f3}(c), respectively, where each point on the plots corresponds to a different value of $t$. These figures demonstrate that $\tau^\ast$ and $f(0)^\ast$ approach $\tau$ and, respectively, $f(0)$ 
almost  
linearly with $1/s^\ast$. 
%%For each value of $t$, we found 
%%the position $s^*$ of 
%%, the corresponding 
%%solid 
%%curve 
%%the minimum of 
%%the function  
%%$\tau(s_{0})$, see Fig.~\ref{f2}, 
%%and analyzed how $1/s^*$  
%%we studied 
%%how the position of the minimum, $1/s^*$, of the solid curve $\tau(1/s_{0})$ in Fig.~\ref{f2} 
%%varies with 
%%depends on 
%%relates to the 
%%corresponding 
%%values 
%%$t^\ast$, 
%%$t$, $\tau^\ast$, and $f^\ast(0) \equiv 
%%P(s^\ast,t^\ast)
%%P(s^*,t)
%%{s^\ast}^{\tau-1}$, corresponding to this minimum. Fig.~\ref{f3} demonstrates that 
%%$t^\ast$ 
%%$t$ approaches $t_c$ 
%%almost  
%%linearly with $1/{s^{\ast}}^{ 2}$, while $\tau^\ast$ and $f(0)^\ast$ approach $\tau$ and, respectively, $f(0)$ 
%%almost linearly with $1/s^\ast$. 
This enables us to make extrapolations to $s^\ast\to\infty$ (the maximum number of equations which we used, $s_{\text{max}}$, was $10^5$) and obtain $t_c$, $\tau$, and $f(0)$ with very high precision. 
%%the precise values. 

%%****
%%To find the precise values of $t_c$, $\tau$, and $f(0)$ we analyzed how the position of the minimum of the solid curve $\tau(1/s)$ varies with the corresponding values of $t\equiv t^*$, $\tau(1/s^*)\equiv \tau^*$, and $P(s^*,t^*){s^*}^{\tau^*-1}\equiv f(0)^*$. 
%%Figure 1.3 demonstrates that $t^*$ approaches $t_c$ linearly with ${1/s^*}^2$, and $\tau^*$, and $f(0)^*$ approach $\tau$ and $f(0)$ linearly with $1/s^*$. This enables us...
%%****

\begin{figure}[t]
%%\sidecaption
% For example, with the graphicx style use
%%\begin{center}
\hspace{4.5pt} \includegraphics[scale=.35]{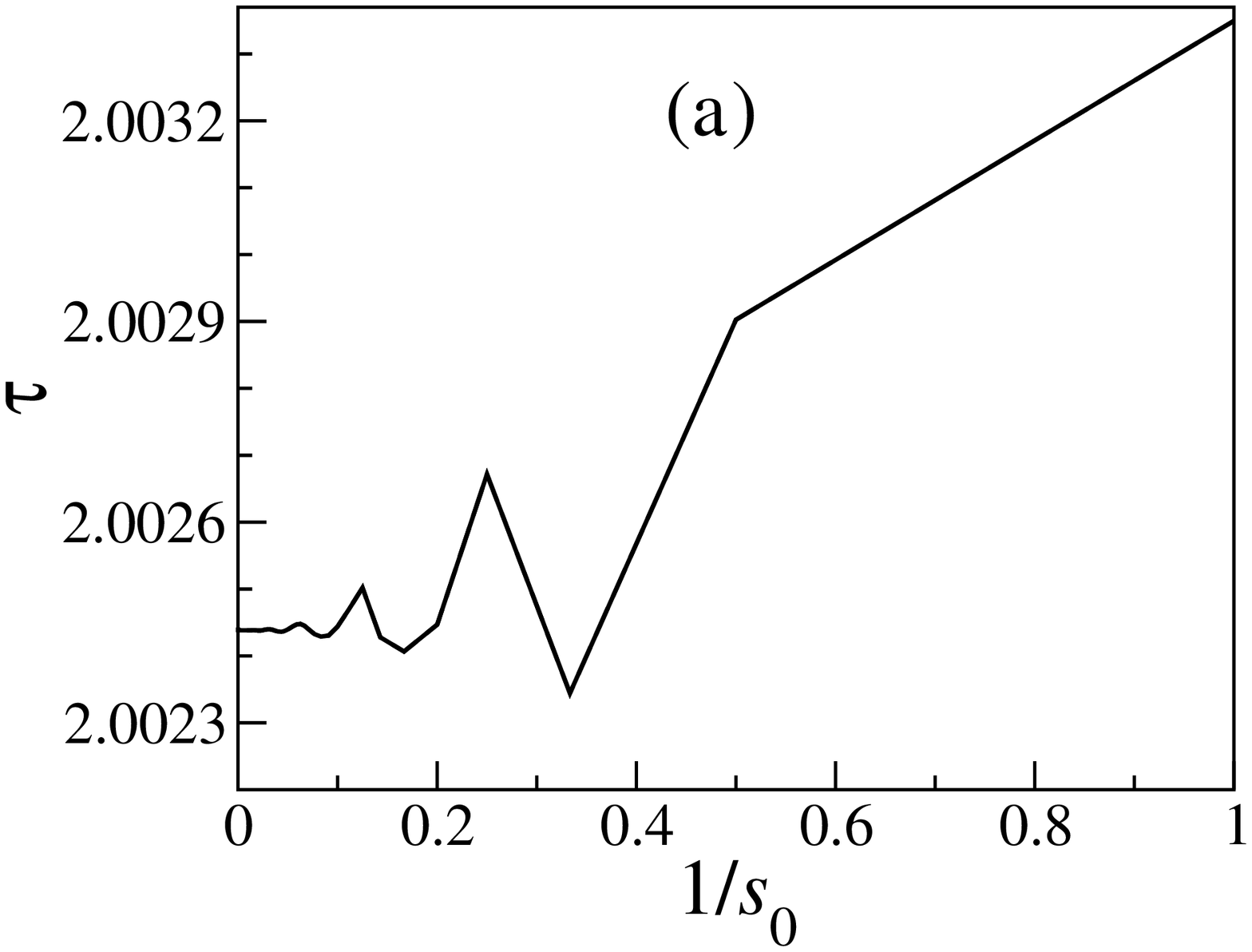}
\\[0pt]
\includegraphics[scale=.35]{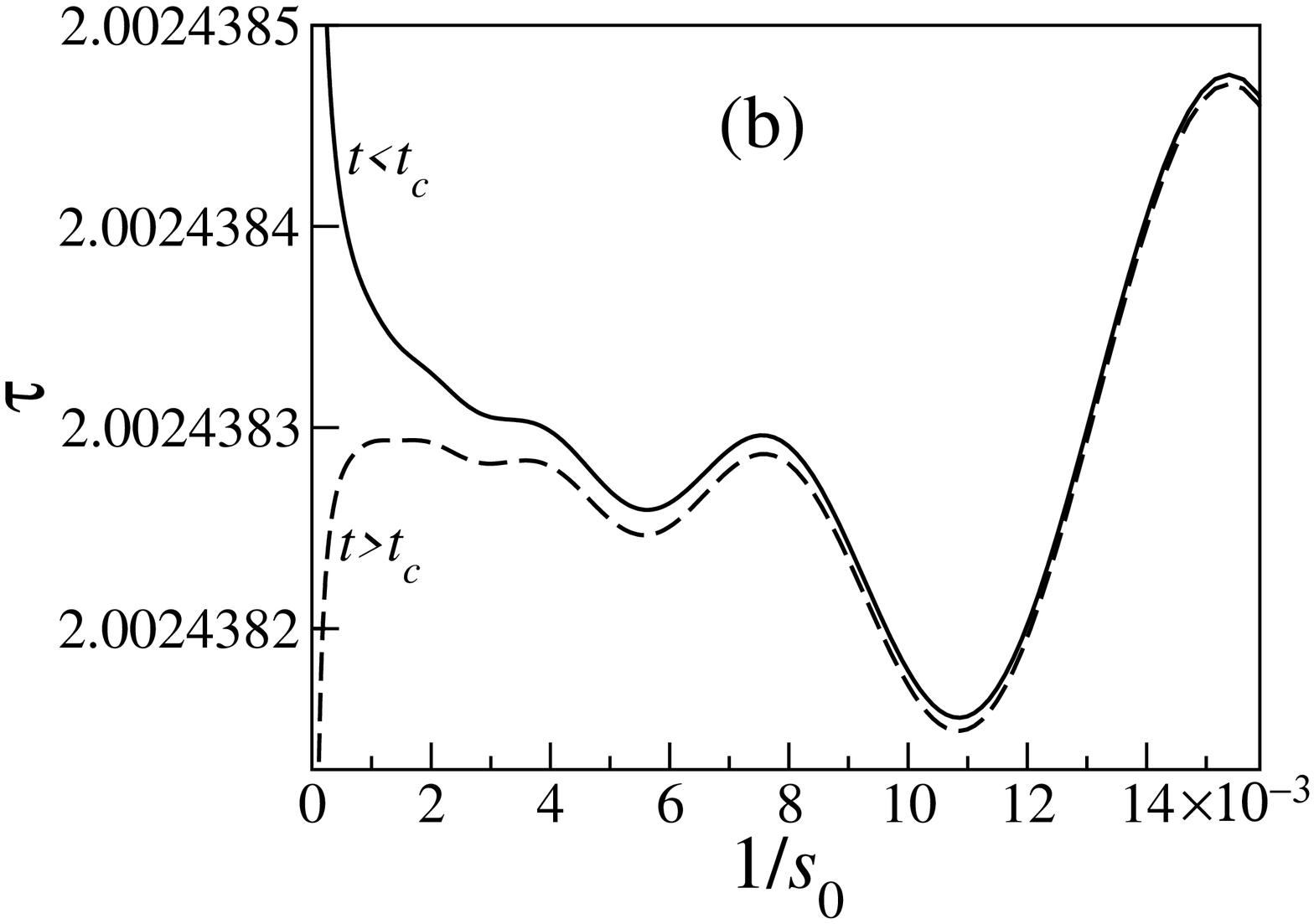}
%%\\[-3pt]
%%\end{center}
%
% If no graphics program available, insert a blank space i.e. use
%\picplace{5cm}{2cm} % Give the correct figure height and width in cm
%
\caption{
Variation of $\tau$, calculated from Eq.~(\ref{eq:norm_cond_2_CEPT}),  with $1/s_{0}$ 
%%, where $s$ denotes the maximum cluster size found by solving numerically the evolution equation, 
in the case of $m=4$. The solid and dashed curves in (a) and (b) correspond to two values of $t$, namely, $t=0.994973560<t_c$  and $t=0.994973564>t_c$, respectively. Panel (a) shows the full range of $s_0$ from $1$ to $s_{\text{max}}=10^5$, and panel (b) shows the part of the same plot in the range of $s_0$ from $1/16\times10^3$ to $s_{\text{max}}=10^5$. In panel (a), the solid and dashed curves are indistinguishable. 
}
\label{f4}       % Give a unique label
\end{figure}

One can even avoid the extrapolation procedure, which may prove to be difficult at $m=4$ and higher. The problem is that for higher $m$ the curves $\tau(s_{0})$ 
%%$\tau(1/s_{0})$ 
oscillate (see Fig.~\ref{f4}), because the distribution $P(s,t)$ oscillates in the range of 
%%sufficiently 
low $s$. The reason for these oscillations is the following. If $m$ tends to infinity, then according to our rules, two smallest clusters in the system 
%%should 
merge at each step together. 
Consequently,  
%%This is why 
single vertices initially merge together into the clusters of size $2$, then these clusters merge into the clusters of $4$ vertices, and so on. This results in the peaks of the distribution at $s=2,4,8,\dots$, which are seen already at $m=4$. 
%%In turn, these peaks lead to the oscillations in Fig.~\ref{f4}. 
Fortunately, the amplitude of the oscillations in Fig.~\ref{f4} decreases with decreasing $1/s_{0}$. 
This enables us to investigate 
%%So one can easily study 
the run away of the curves from the precise value of $\tau$ at small $1/s_{0}$. 
As $t$ approaches $t_c$ from below or above, the run away in direction of infinity or $2$, respectively, occurs at smaller and smaller values of $1/s_{0}$. 
%%********
%%We adjust $t$ progressively 
%%%%and find $t_c$, with high precision, by adjusting $t$ 
%%in such a way that the run away takes place at the smallest value of $1/s_{0}$,
%%which is $10^{-5}$ for the results presented in this paper. 
%%Even by visual inspection we can easily 
It is easy to identify an interval where $t_c$ should lie. 
The lower bound of this interval is the biggest value of $t$ for which the curve $\tau(1/s_{0})$ still shows a 
%%clear 
trend to infinity at the smallest $1/s_{0}$, i.e., $1/s_{\text{max}}$. 
The higher bound of the interval is the smallest value of $t$ that corresponds to a curve still demonstrating a trend to $2$ at $1/s_0 = 1/s_{\text{max}}$.
%Precise values 
Corresponding intervals for $\tau$ and $f(0)$ are then obtained by using Eqs.~(\ref{90}) and (\ref{eq:norm_cond_2_CEPT}). 
We adjust $t$ progressively, using the shooting method, 
%%and find $t_c$, with high precision, by adjusting $t$ 
in such a way that the run away takes place at the smallest value of $1/s_{0}$. 
%%, which is $10^{-5}$ for the results presented in this paper. 
%%latest (i.e., occurring at the smallest value of $1/s_{0}$) run away.        
%%The 
%%length of the 
%%intervals calculated in this way 
%%is 
%%are remarkably small. 
Even using only the numerical solution of the first $10$ evolution equations 
%(i.e., for a smallest $1/s_{0}$ of $0.1$) 
this method gives $t_c=0.924(2)$ and $\tau = 2.047(3)$ for $m=2$.
The 
%%high 
precision of all the results shown in Table~\ref{table1} was attained by direct 
%%visual 
inspection of the curves $\tau(1/s_{0})$, for a $s_{0}\leq s_{\text{max}} =10^{5}$. 
We tested our method on ordinary percolation ($m=1$) started from isolated nodes. Using $s_{\text{max}}=1000$ recovers the exact values $t_c=1/2$  and $\tau=5/2$ with precision $2{\times} 10^{-5}$ and $8{\times} 10^{-4}$, respectively. 

%The method proposed~\cite{Vijayaraghavan:vnwd13} can't produce more precise results is likely related due to the fact that the scaling form of the cluster

%%%%%%%%%%%%%%%%%%%%%%%%%%%%%%%%%%%%%%%%%%%%%%
%%%%%%%%%%%%%%%%%%%%%%%%%%%%%%%%%%%%%%%%%%%%%%

%%
%%

%%%%%%%%%%%%%%%%%%%%%%%%%%%%%%%%%%%%%%%%%%%%%%
%%%%%%%%%%%%%%%%%%%%%%%%%%%%%%%%%%%%%%%%%%%%%%
 
 Another approach for a model of this class was used in Ref.~\cite{Vijayaraghavan:vnwd13} to estimate the percolation threshold position imposing the strong assumption that the cluster size distribution $P(s,t)\propto s^{1-\tau}e^{-cs}$, where $c$ is time-dependent, and $c(t_c)=0$. 
%%In Ref.~\cite{Vijayaraghavan:vnwd13} it was proposed a method for the estimation of the percolation threshold that is based on the scaling properties of the cluster size distribution as well. There it is assumed that $P(s,t)\propto s^{1-\tau}e^{-cs}$, where $c$ is a function of time which should become zero at $t_c$, leading to a pure power-law critical distribution. 
In this way, after solving $10^5$ evolution equations, they achieved the same precision of $t_c$ (or even worse) which our method provides with only $10$ equations. 
%% However, the fitting of data gives $c\sim0$ on an interval of size roughly as large as the interval given by our method with only $10$ evolution equation. 
One should emphasize that 
%%This behavior of the fitting parameter $c$ is likely to be related to the fact that, contrarily to what was assumed, the cut-off function of 
the actual scaling form of the cluster size distribution 
%%is not 
essentially deviates from a simple exponential, see Ref.~\cite{daCosta:ddgm10}. 
%%,daCosta:ddgm}. 

%%%%%%%%%%%%%%%%%%%%%%%%%%%%%%%%%%%%%%%%%%%%%%
%%%%%%%%%%%%%%%%%%%%%%%%%%%%%%%%%%%%%%%%%%%%%%

\begin{table*}
\caption{Characteristics of the standard percolation ($m=1$) and explosive percolation ($m=2,3,4$) transitions.} 
%%$t_c$ is the critical point, $\beta$ and $\tau$ are the critical exponents of the percolation cluster and the finite cluster size zistribution, respectively, $f(0)$ is the critical amplitude, and shows}
%%\begin{center}
\begin{tabular}{ccccc}
\hline
\noalign{\smallskip}
%%& & & & 
%%\\
%%[-9pt]
$m$     &   1   &  2  &  3  &  4 
\\[3pt]
\hline
%%\noalign{\smallskip}
%%\svhline
%%\noalign{\smallskip}
%%\hline
%%[0pt] 
%%& & & & 
\\[-8pt]
%%%%%\hline
%%%%%&&&&
%%%%%\\
$t_c$   & $1/2$ &$\ $  
%%0.92320750930(2) 
0.923207509297(2)
$\ $  & 
%%0.98179531735(5) 
0.9817953173509(2)
& 
%%0.99497356260(5)
0.99497356260563(2)  
\\
$\beta$ &   1   &  
%%0.05557106(3) 
0.05557108(1)
& 
%%0.01042872(2) 
0.010428725(1)
&  
%%0.00248067(1) 
0.0024806707(2)
\\
$\tau$  & $5/2$ & 
%%2.04763044(2) 
2.04763045(1)
& 
%%2.00991188(2)  
2.009911883(1)
&  
%%2.00243833(1) 
2.0024383299(1)
\\
$d_{\text{uc}}$  & $6$ & 
2.4445686(1)
& 
2.12514470(2)
&  
2.039690731(3)
\\
$f(0)$  
&  $1/\sqrt{2\pi}{\approx}0.3989$   
&  
%%0.04619068(2)  
0.04619071(1)
& 
%%0.00983139(1) 
0.009831398(1)
&  
%%0.00244(2)  
0.0024320386(1) 
\\
$P(
%%s{=}
1,t_c)$
& $1/e{\approx}0.3678$ &  
%%0.04859280(1)  
0.0485928295546(4)
&
%%$\ $  0.011721464802(3)$\ $  
0.01172146480245(2)
& 
%%$\ $ 0.003343067143(3)
%%0.0033430671431325(5)
0.003343067143133(1)
\\
\noalign{\smallskip}\hline\noalign{\smallskip}
%%[2pt]
%%\hline
\end{tabular}
%%\end{center}
\label{table1}
\end{table*}

%%%%%%%%%%%%%%%%%%%%%%%%%%%%%%%%%%%%%%%%%%%%%%
%%%%%%%%%%%%%%%%%%%%%%%%%%%%%%%%%%%%%%%%%%%%%%

%%%%%%%%%%%%%%%
%%%%%%%%%%%%%%%%%%%%%%%%%%%%%%%%%%%
%%%%%%%%%%%%%%
%%%%%%%%%%%%%%%%%%%%%%%%%%%%%%%%%%%

%
\section{Critical exponents and amplitudes}
\label{sec:4}   

The results of the application of this numerical method to the models with $m=2$, $3$, and $4$ are presented in Table~\ref{table1}. For comparison, in the first column of the table, we show the exact values for the ordinary  percolation problem ($m=1$). The values of the exponent $\beta$ 
and the upper critical dimension $d_{\text{uc}}$
are found 
%%from $\tau$ 
by using relations~(\ref{e4}) and (\ref{e4a}). 
In the case of $m=2$, the values presented in the table agree with our results \cite{daCosta:ddgm10}, although the precision of the numbers obtained in the present work is much higher despite the fact that here we solved $10$ times fewer evolution equations than in Ref.~\cite{daCosta:ddgm10}. 
Furthermore, the results in the table for the models with $2\leq m\leq 4$ agree with those obtained from equations for scaling functions (we will consider this alternative method elsewhere). 
%%\cite{daCosta:ddgm}). The table shows 
%%that as is natural, 
As $m$ increases, the difference $1-t_c$ decreases, and the exponent $\beta$ of the giant component size also decreases. The critical amplitude $f(0)$ is close to $P(1,t_c)$, especially when $m \leq 3$. This closeness indicates that the deviations from a power-law asymptotics in these problems are small even at low values of $s$. Note that $f(0)>P(1,t_c)$ for classical percolation, while the opposite is true for the explosive percolation transition. The values of $\beta$ are remarkably small. In particular, in the case of $m=4$, $\beta$ is close to $1/400$. 
This produces an extremely ``sharp'' 
%%continuous 
transition whose continuity is virtually unobservable even in astronomically large though finite systems. 
%%While simulating these systems, the smallest step in $t$ corresponds to addition a single link and equals $1/N$, that is the inverse size of a system. Let $m$ be $4$. Then the minimum jump at $t_c$, which one can observe simulating even an unrealistically large system of $10^{400}$ nodes, is of the order of $(10^{-400})^{1/400} = 10^{-1}$, which makes observation of a continuous transition in simulations impossible. We suggest that finite size effects in this situation hardly can be investigated. Even in the case of $m=2$, in which $\beta$ is close to $1/18$, simulating a system of $10^{18}$, one cannot observe a jump smaller than of the order of $10^{-1}$ at the critical point.  
One should note that our way to vary exponents in these non-universal systems by changing $m$ is not unique. 
For example, other explosive percolation models, introduced in Refs.~\cite{Fan:fllc12,Andrade:ah13}, showed  
%%similar behavior 
a decay of $\beta$, controlled by a different, specially introduced model parameter. 
%%which diminishes with variation of a .........
%%also approaches $0$ as a model parameter is varied. .........

%%%%%%%%%%%%%%%
%%%%%%%%%%
%%%%%%%%%%%%%%
%%%%%%%%%%%

\section{Discussions and conclusions}
%\label{sec:5}

Our results show a rapid decrease of the critical exponent $\beta$ values with increasing $m$. For $m=4$, exponent $\beta$ is about 20 times smaller than already its tiny value for $m=2$. 
This indicates that exploration of this transition by means of numerical simulations at higher $m$ is hardly possible. 
Table~\ref{table1} demonstrates that the upper critical dimension $d_{\text{uc}}$ quickly approaches $2$ with increasing $m$. So the explosive percolation transition in models of this class placed on two-dimensional lattices is very close to its upper critical dimension. 
This suggests that the critical characteristics of explosive percolation on two-dimensional systems should be close to what was obtained in this paper.  

We emphasize that our results were found for models in which evolution is determined by purely local optimization rules. It means that each new interconnection uses only a finite amount of information. 
%%You do not need to know about all clusters in the system including the giant one, 
In particular, to establish a new link, we do not need to know which of clusters is the biggest in the system.  (Indeed, to find the largest cluster, one has to know about all of them.) It is the local optimization rule that leads to continuity of the explosive percolation transition in these models. In more exotic models employing various global optimization algorithms and their variations, discontinuities may occur \cite{Schrenk:sah11,Chen:cczcdn13,Riordan:rw12,Chen:cncjszd13,Schrenk:sfdadh12,Araujo:ah10a,Cho:chhk13}. 

In summary, we have proposed an effective numerical method enabling us to find characteristics of explosive percolation transitions with high precision. We obtained the critical points,  critical exponents, and critical amplitudes 
%%of scaling functions of these transitions 
in a set of representative models. The fact that critical exponents are model dependent demonstrates the non-universality of critical phenomena for this phase transition. 
%%of explosive percolation. 
Our results confirm the conclusion that explosive percolation transitions are continuous, 
%%namely 
with a power-law form of the cluster size distribution at the critical point. 
%%namely, the assumption of a critical power-law cluster size distribution. This kind of asymptotic behavior is observed for all the models ($m=2,3,4$). 
Based on our observations, we suggest that 
%% Moreover, we conclude that 
in a wide range of models of this kind 
%%with rules similar to those described in sections~\ref{sec:1} and \ref{sec:2}, 
the explosive percolation transition is continuous, including in particular, the models considered in Refs.~\cite{Achlioptas:ads09,Nagler:nlt11,Ziff:z09,Radicchi:rf09,Cho:ckp09,Cho:ckk10,D'Souza:dm10,Manna:mc11,Friedman:fl09,Choi:cyk11} . 
 
Our approach provides a useful tool for a quantitative description of a new class of critical phenomena in non-equilibrium systems and irreversible processes. 
%%Due to the absence of universality, the critical exponents of these transitions depend on the specific model. 
Moreover, the applicability of this numerical method is not limited to explosive percolation models. 
%%We notice that, 
Since the method relies only on generic scaling properties, it is suitable to a wide range of continuous phase transitions in non-equilibrium systems.

%%%%%%%%%%%%%%%%%%%%%%%%%%%%%%%%%%%%%%%%%%%%%%%%%%%%%%%%%%%%%%%%%%
%%%%%%%%%%%%%%%%%%%%%%%%%%%%%%%%%%%%%%%%%%%%%%%%%%%%%%%%%%%%%%%%%%

\begin{acknowledgments}
This work was partially supported by the FCT
Project PTDC/MAT/114515/2009 and FET IP Project MULTIPLEX 317532.
%%, and also by the SOCIALNETS EU project.
\end{acknowledgments}
%%%
%%%

\end{document}